\documentclass[aip,jcp,reprint]{revtex4-1}

\usepackage{amsmath,amssymb}
\usepackage{graphicx}% Include figure files
\usepackage{dcolumn}% Align table columns on decimal point
\usepackage{bm}% bold math
\usepackage{color}

\usepackage[utf8]{inputenc}
\usepackage[T1]{fontenc}
\usepackage{mathptmx}
\usepackage{braket}
\usepackage[english]{babel}
\usepackage[nolist]{acronym}
\usepackage{scrextend} % To reference some of the footnotes in tables for the second time

\newcommand{\FeWat}{[Fe(H$_2$O)$_6$]$^{2+}$}
\newcommand{\FeWatAm}{[Fe(H$_2$O)$_5$(NH$_3$)]$^{2+}$}
\newcommand{\FeAm}{[Fe(NH$_3$)$_6$]$^{2+}$}
\newcommand{\FeCN}{[Fe(H$_2$O)$_5$(CN)]$^{+}$}
\newcommand{\TiO}{[TiO$_6$]$^{8-}$}
\newcommand{\CrWat}{[Cr(H$_2$O)$_6$]$^{3+}$}
\newcommand{\NiWat}{[Ni(H$_2$O)$_6$]$^{2+}$}
\newcommand{\FeCO}{[Fe(CO)$_5$]$^{0}$}

\newcommand{\Ssq}{$\langle \hat S^2 \rangle$}

\newcommand{\Supp}{Supplement}

\draft % marks overfull lines with a black rule on the right

\begin{document}

%\title[Ultrafast electron dynamics]{Effect of the \rt{nature of central atom} chemical environment on ultrafast electron dynamics in \rt{transition metal complexes} triggered by \rt{soft X-ray} light} %Title of paper
% Conditions for effective spin-flip
% Effect of the chemical environment on spin-orbit coupling driven spin crossover
\title[Ultrafast spin dynamics]{Effect of chemical structure on the ultrafast spin dynamics in core-excited states}

\author{Vladislav Kochetov}
%\email[]{vladislav.kochetov@uni-rostock.de}
%\thanks{}
\affiliation{Institut f\"{u}r Physik, Universit\"{a}t Rostock, A.-Einstein-Strasse 23-24, 18059 Rostock, Germany}

\author{Huihui Wang}
%\thanks{}
\affiliation{State Key Laboratory of Quantum Optics and Quantum Optics Devices, Institute of Laser Spectroscopy, Shanxi University, 030006, China}

\author{Sergey I. Bokarev}
\email[]{sergey.bokarev@uni-rostock.de}
%\thanks{}
\affiliation{Institut f\"{u}r Physik, Universit\"{a}t Rostock, A.-Einstein-Strasse 23-24, 18059 Rostock, Germany}

\date{\today}

\begin{abstract}
		Recent developments of the sources of intense and ultrashort X-ray pulses stimulate theoretical studies of phenomena occurring on ultrafast timescales. 
		In the present study, spin-flip dynamics in transition metal complexes triggered by sub-femtosecond X-ray pulses are addressed theoretically using a density matrix-based time-dependent configuration interaction approach.
	    The influence of different central metal ions and ligands on the character and efficiency of spin-flip dynamics is put in focus.
	    According to our results, slight variations in the coordination sphere do not lead to qualitative differences in dynamics, whereas the nature of the central ion is more critical.
	    However, the behavior in a row of transition metals demonstrates trends that are not consistent with general expectations.
	    Thus, the peculiarities of spin dynamics have to be analyzed on a case-to-case basis.
\end{abstract}


\maketitle

\begin{acronym}
	\acro{AO}{Atomic Orbital}
	\acro{CAS}{Complete Active Space}
	\acro{CASPT2}{Complete Active Space Second Order Perturbation Theory}
	\acro{CI}{Configuration Interaction}
	\acro{CSF}{Configuration State Function}
	\acro{HF}{Hartree-Fock}
	\acro{HHG}{High Harmonics Generation}
	\acro{MCSCF}{Multi-configurational Self-Consistent Field}
	\acro{MO}{Molecular Orbital}
	\acro{RAS}{Restricted Active Space}
	\acro{RASPT2}{Restricted Active Space Second Order Perturbation Theory}
	\acro{RASSCF}{Restricted Active Space Self-Consistent Field}
	\acro{RASSI}{Restricted Active Space State Interaction}
	\acro{RIXS}{Resonant Inelastic X-ray Scattering}
	\acro{SCF}{Self-Consistent Field}
	\acro{SF}{Spin--Free}
	\acro{SO}{Spin--Orbit}
	\acro{SOC}{Spin--Orbit Coupling}
	\acro{TD-RASCI}{Time-Dependent Restricted Active Space Configuration Interaction}
	\acro{XAS}{X-ray Absorption Spectrum}
	\acro{XES}{X-ray Emission Spectrum}
	\acro{XFEL}{X-ray Free Electron Laser}
	
\end{acronym}

%TODO write everywhere nhbar instead of n for the values of S^2
\section{Introduction} \label{sec:intro}

	Novel light sources such as \ac{HHG} and \ac{XFEL} are steadily improving in terms of increasing intensity, energy and shortening pulse duration and temporal resolution down to attoseconds.\cite{Hentschel_N_2001,Kienberger_N_2004,Grguras_NP_2012,Gaumnitz_OE_2017, Maroju_N_2020}
	Such an advance allows one to study electron dynamics on a few femtosecond and subfemtosecond timescales.~\cite{Schultz_book_2014, Young_JPB_2018} 
	The key point is the preparation of a superposition of quantum states by pulses, which have a broad linewidth in the frequency domain. 
	This non-stationary superposition then coherently evolves in time. 
	Examples of such behavior were demonstrated experimentally and reinforced theoretically for the different cases of charge migration. \cite{Worner_SD_2017} 
	Due to their ultrafast character, the early electron dynamics appear to be almost isolated from nuclear motion and other effects taking place at longer times.
	
		\begin{figure}[b]
		\includegraphics[width=1\columnwidth]{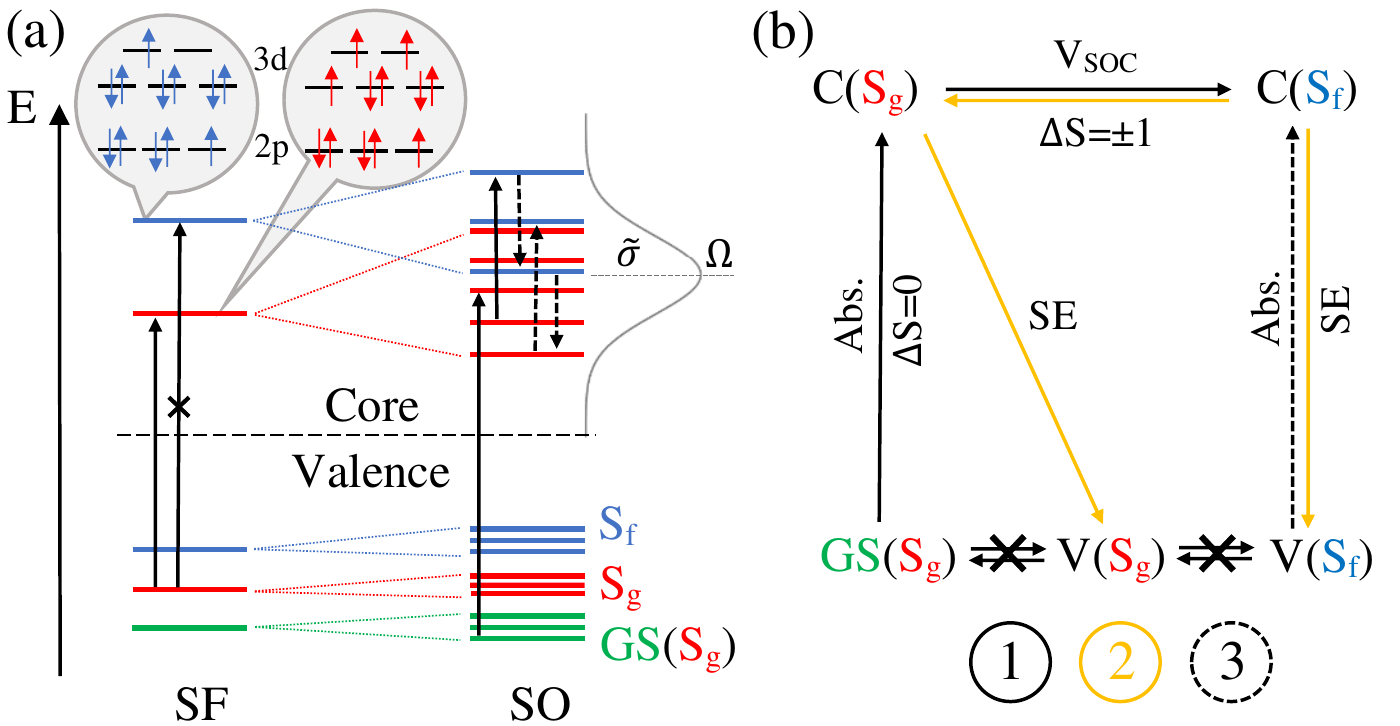}
		\caption{\label{fig:scheme} 
			(a) Scheme of many-electron energy levels in the system without (left) and with (right) \ac{SOC}. 
			States of different spin are marked with red and blue colors; we additionally distinguish the ``ground'' states (green), which can be populated due to the finite temperature.
			The light pulse with carrier frequency $\Omega$ and bandwidth $\tilde \sigma$ is shown in gray; it prepares the superposition of the \ac{SOC} states.
			(b) Population pattern of the core (C) and valence (V) states with spins $S_g$ and $S_f$ enabled by the light absorption (Abs.), stimulated emission (SE), and \ac{SOC} ($V_{\rm SOC}$).
			The style of arrows corresponds to the primary, secondary, and tertiary processes.
		}
	\end{figure}
	
	Another kind of coherent dynamics reported recently and considered in this paper is the spin dynamics initiated by X-ray light.~\cite{Wang_PRL_2017, Wang_PRA_2018} 
	Its principle is briefly illustrated in Fig.~\ref{fig:scheme}(a) showing the many-body state patterns without (\ac{SF}) and with (\ac{SO}) strong \acf{SOC}, which is characteristic for core-excited states.
	Initially, only the spin-allowed transitions with $\Delta S=0$, i.e., between green and red \ac{SF} states, are occurring upon light absorption.
	It was shown~\cite{Wang_MP_2017, Wang_PRA_2018} that the creation of a core-hole in $2p$ orbitals, i.e., the L$_{2,3}$-edge absorption, in transition metal complexes is followed by the mixing of states with different spins evolving in time. 
	For certain pulse characteristics, this process leads to a spin-flip, taking place within about a femtosecond, which is extremely fast compared to the conventional spin-crossover times, taking, as a rule, more than 50~fs.~\cite{Hauser_CCR_1991,Forster_CCR_2006a,Marian_WCMS_2012} 
	However, in exceptional cases, it may take notably less time.~\cite{Mai_CS_2019} 
	Since \ac{SOC} for the deeper holes with non-zero angular momentum is in general much larger than in the valence band, simulating dynamics initiated by high-energy photons is of interest.
	
	From the general viewpoint, the ultrafast spin-flip should occur when a superposition of states with $2p_{3/2}$ and $2p_{1/2}$ core holes is effectively prepared by excitation with a broadband pulse. 
	Thus, one can expect this process to be purely dictated by the properties of these core holes, with the chemical environment and the details of the pulse characteristics being less relevant.
	However, previous works \cite{Wang_PRA_2018, Wang_PRL_2017} concluded that the carrier frequency and the width of the pulse are essential to trigger the efficient spin-flip transition.
	In particular, the creation of superposition of the $2p_{3/2}$ and $2p_{1/2}$ core holes was not always a prerequisite for such a spin transition to occur.
	Moreover, the peculiarities of the dynamics were discussed only for one particular system - hexaaqua iron (II) complex \FeWat, where a sub-femtosecond transition from quintet to triplet states has been observed. 
	Therefore, the question of the conditions for the efficient spin transition calls for additional study.
	
	The central question of this article is how the nature of the excited  metal atom and its chemical environment (coordination sphere) influences the dynamics and the spin-flip yield. 
	Focusing on the transition metal (Ti, Cr, Fe, Ni) complexes with different weak- and strong-field ligands, we deduce the compounds, from which one can expect significant changes in the populations of states with different spins, and discuss conditions, at which one can observe them.

	The article is organized as follows: First, we present the theoretical method used in this work in Section~\ref{sec:method}.
	Further, the logic behind the choice of the objects under investigation is explained in Section~\ref{sec:species} and the essential parameters of the computation are presented in Section~\ref{sec:comp_details}.
	The influence of ligands and the nature of the central metal on the dynamics are presented in Section~\ref{sec:results} and are further analyzed in Section~\ref{sec:discussion}.
	Finally, the conclusions are given in Section~\ref{sec:conclusions}.

\section{Method}\label{sec:method}
	The approach we use for the study of the populations of spin states is the quite general density matrix-based \ac{TD-RASCI} described elsewhere.~\cite{Wang_MP_2017,Tremblay_JCP_2008} 
	Since the whole process of interest lasts no more than few fs, nuclear motion is neglected in the calculation. 
	The dynamics of an open system are described by the Liouville-von Neumann equation~\cite{May2011} 
	\begin{equation}\label{eq:liouville}
	\frac{\partial}{\partial{t}}\hat{\rho}=-i[\hat{H},\hat{\rho}]+\mathcal{D}\hat{\rho} \, .
	\end{equation}
	If one writes the Hamilton operator in the basis of configuration state functions,~\cite{Szabo1996} it has the form
	\begin{equation} \label{eq:hamiltonian}
		\mathbf{H}(t)= \mathbf{H}_{\rm CI}+\mathbf{V}_{\rm SOC}+\mathbf{U}_{\rm ext}(t) \, .
	\end{equation}
	Here, $\mathbf{H}_{\rm CI}$ and $\mathbf{V}_{\rm SOC}$ are the \ac{CI} Hamiltonian, responsible for electron correlation effects, and the \ac{SO} interaction part, respectively.
	The light field contributes to the Hamiltonian with the time-dependent light-matter interaction term 	$\mathbf{U}_{\rm ext}(t) = -{\mathbf{d}} \cdot \vec{E}(t)$ in the dipole approximation, where $\mathbf{d}$ is a transition dipole matrix and $\vec{E}(t)$ is an external electric field.	
	
	However, in practical applications to the spin-flip dynamics, the so-called	\ac{SF} basis $\{\Phi_{i}^{(S_i,M_{S_i})}\}$ is more convenient.
	These $\{\Phi_{i}^{(S_i, M_{S_i})}\}$ are the eigenfunctions of $\mathbf{H}_{\rm CI}$ and correspond to a particular spin $S$ and its projection $M_S$ onto the quantization axis and thus are the eigenfunctions of the $\hat S^2$ and $\hat S_z$ operators.
	The term $\mathbf{V}_{\rm SOC}$ couples different \ac{SF} functions by \ac{SOC} such that the eigenfunctions of $\mathbf{H}_{\rm CI}+\mathbf{V}_{\rm SOC}$ are the linear combinations of \acp{SF} with different spin.
	 The density matrix in the \ac{SF} basis is given by
	\begin{equation}\label{eq:rho_sf_basis}
	\pmb{\rho} (t) = \sum_{i,j} 
	           { \rho_{ij}^{(S_i,M_{S_i}),(S_j,M_{S_j})}(t) \ket{\Phi_i^{(S_i,M_{S_i})}}\bra{\Phi_{j}^{(S_j,M_{S_j})}}} \, .
	\end{equation}

	Diagonal elements $\rho_{ii}$ in such representation are the populations of the corresponding \ac{SF} states $\ket{\Phi_i^{(S_i,M_{S_i})}}$.
	For the simplicity of analysis, the values of state populations $\rho_{ii}$ with the same total spin $S$ have been summed  
	\begin{equation}\label{eq:populations}
		P({S}) = \sum_{\substack{i \\ S_i=S}} 
		 \rho_{ii}^{(S_i, M_{S_i}),(S_i, M_{S_i})} 
	    \,. 
	\end{equation}
	We use the following notation for the different groups of states: $P({GS})$ is the population of a single or several ground states with the spin $S_g$ split by \ac{SOC} and found in thermal equilibrium at finite temperature. 
	$P({S_g})$ is the population of the excited states with the same spin $S_g$, which is also called the ``main'' spin below.
	$P({S_f})$ is the population of the excited states with the ``flipped'' spin $S_f$ different from that of the ground one.
	The mean value of the spin squared operator has been calculated in \ac{SF} basis as
	{\begin{equation}\label{eq:s_square}
		\langle \hat{S}^2 \rangle = \mathrm{tr}[\hat\rho \hat{S}^2] = \sum_{i} \rho_{ii} S_{ii}^{2} = \sum_{S} {P({S}) \cdot S (S+1)}
	\end{equation}}
	and is used further as an integral characteristic of the spin-flip process.
	%TODO Do we need this note? And the discussion below
	%\bt{One should note that in the time-resolved X-ray absorption spectroscopy,~\cite{Bressler_S_2009} X-ray emission spectroscopy or resonant inelastic X-ray scattering~\cite{Wernet_Nat_2015,Wernet_PTRSMPES_2019} experiments, aiming at tracing the spin dynamics, the populations of the states rather than the $\langle \hat S_z \rangle$ or \Ssq\ values are measured in contrast to magnetic moment measurements.~\cite{Weber_IC_2007,Halcrow2013}
	%In this respect, the time evolution of {$P({S})$ appears} to be a more natural characteristic whilst the value $\langle \hat{S}^2 \rangle$ is a convenient way of representation.
	%Thus, in this article we employ both characteristics.}

	%Below, only subscripts, i.e. $S_g, S_f$, will be used to label the populations, 
	%if used with $GS$ label, 
	%meaning partial populations with corresponding spin.
	
	For simplicity, the incoming electric field was chosen to be a single linearly-polarized pulse with a temporal Gaussian envelope
	\begin{equation}\label{eq:pulse}
	  \vec{E}(t) = A \vec{e}\ \exp{(-{t^2}/({2\sigma^2}))}\sin (\Omega t)\ ,
	\end{equation}
	although the pulse trains are more efficient to induce spin-flip transitions.~\cite{Wang_PRA_2018}
	Here, $A$, $\vec e$, and $\Omega$ are the amplitude, polarization, and carrier frequency.
	The pulse width $\sigma$ was chosen such as to cover a wide range of valence-core excitations; thus, it corresponds to the ultrashort pulse in the time domain.

	Propagation of the density matrix according to Eq.~\eqref{eq:liouville} is performed in \ac{SF} basis with the Runge-Kutta-Cash-Karp method \cite{Cash_ATMS_1990,Press1996} of the 4(5) order of accuracy. 
	In the initial density matrix, the states were populated according to the Boltzmann distribution: $\rho_{ij}(0)=\delta_{ij}\exp(-E_i/(kT))$.
	Neither Auger decay and photoionization nor coupling to the environment were taken into account in the present study for the simplicity of interpretation. 
	Thus, we omit the dissipation operator in Eq.~\eqref{eq:liouville} for the current application, i.e., $\mathcal{D}=0$. 
	However, we note that such effects may matter for a general case.~\cite{Tremblay_JCP_2011, Wang_PRA_2018} 
	All the computations for the density matrix propagation are carried out employing the locally modified version of the \texttt{OpenMOLCAS} package.~\cite{FernandezGalvan_JCTC_2019}

\section{Investigated species} \label{sec:species}

	\begin{table*}
		\caption{\label{tab:objects}Details of the geometric and electronic structure of studied complexes: the metal-ligand distance, the number of $3d$ electrons in the ground state, the magnitude of L$_3$/L$_2$ energy splitting, and the total number of \ac{SOC} electronic states with different multiplicities considered in the dynamics.}
		\begin{ruledtabular}
			\begin{tabular}{llllll}
				
				Compound       & R(M--L)\footnote{Due to the Jahn-Teller effect or the presence of axial/equatorial ligands, distances may vary. All different distances are given in this case.}, \AA{}  & $3d$ electrons & L$_3$/L$_2$ \ac{SOC} splitting\footnote{The energy splitting between the L$_3$ and L$_2$ highest peaks.}, eV & States ($2S_g+1$)  & States ($2S_f+1$) \\	
				\hline
				
				\textit{Reference} & & & & & \\
				\FeWat         & 2.04, 2.27    & 6 & 12.8 & 175 (5) & 585 (3) \\
				\hline
				
				\textit{Set 1} & & & & & \\
				\FeWatAm       & 2.16, 2.20, 2.26\footnote{\label{note:axial}The order of distances: {equatorial H$_2$O, axial H$_2$O, axial NH$_3$ or CN$^-$ ligand}.}   &   & 12.6 &  &  \\	
				\FeAm          & 2.30             & 6 & 12.6 & 175 (5) & 585 (3) \\
				\FeCN          & 2.05, 2.11, 1.92\footref{note:axial}  &   & 12.5 &  &  \\
				
				\hline
				
				\FeCO          & 1.68, 1.77        & 8 & 11.0 & 751 (1) & 3015 (3) \\
				\hline
				
				\textit{Set 2} & & & & & \\
				\TiO           & 2.02, 2.04, 2.27   & 0 & 5.3 & 16 (1) & 45 (3) \\
				\CrWat         & 2.00 (1.97)\footnote{The distance optimized with the \ac{CASPT2}.}  & 3 & 7.2 & 640  			(4) & 650 (2), 90 (6) \\
				\NiWat         & 2.09            & 8 & 17.9 & 75 (3) & 30 (1)

			\end{tabular}
		\end{ruledtabular}
	\end{table*}

	In this study, we have focused on transition metal complexes as convenient objects to study the effect of chemical structure on the spin dynamics.
	These complexes exhibit states of different multiplicities that can be close in energy, as shown in Fig.~\ref{fig:scheme}. 
	Their relative energies are governed by the interplay of the ligand field splitting and pairing (exchange) energy and depend on 	the position of the ligand in the spectrochemical series. 
	In turn, \ac{SOC} increases from left to right in the row of transition metals. 
	Both effects influence relative stability and spin crossover properties.~\cite{Hauser_CCR_1991, Forster_CCR_2006a}

	When talking about the soft X-ray excitation of metal atoms, the electronic states relevant for such dynamics are of the $2p \rightarrow 3d$ type and are strongly dipole allowed, in contrast to the weaker $2p \rightarrow 4s$ ones. 
	Depending on the number of electrons in the $d$-shell, the number of these states also varies because of the difference in the available $d$-holes for an excited $2p$ electron. 
	Variation in the total number of accessible states may also strongly influence the spin dynamics, in addition to differences in \ac{SOC} strength, and its impact has been studied here.
	Further, the $3p \rightarrow 3d$ transitions could also be relevant for observing spin-dynamics and seem to be more attractive as M-edge absorption requires less energetic radiation, but at the same time, \ac{SOC} for the $3p$ holes is notably smaller than for the $2p$ ones.
	This type of transitions has also been considered in this paper.
	
	Two main sets of compounds were studied, see Table~\ref{tab:objects}. 
	The weak-field $d^6$ iron hexaaqua complex \FeWat with a quintet ground state is used as a reference as it has been recently studied and demonstrated an efficient spin-flip transition.~\cite{Wang_PRA_2018} 
	The \textit{set 1} includes this hexaaqua iron (II) complex and its derivatives with the general formula $\mathrm{[FeX}_n \mathrm{(H_2O)}_{6-n}\mathrm{]}^{2+}$, where water molecules are partially or completely replaced by stronger ligands X=NH$_3$ ($n=1$ or 6), or even stronger CN$^-$ ($n=1$). 
	This set is intended to test the influence of ligand strength; complexes of \textit{set 1} are listed in Table~\ref{tab:objects} in the ascending order of the spectrochemical strength of ligands. 
	 
	The second set comprises six-coordinated complexes \TiO\ and [M(H$_2$O)$_6$]$^{n+}$ with M=Cr, Fe, Ni ordered by the \ac{SOC} value or equivalently by their nuclear charge. 
	The perovskite building block, \TiO cluster, has been chosen because of its high relevance to many functional materials; besides, it resembles the reference \FeWat complex, also having a nearly octahedral coordination sphere of oxygen atoms.
	It is also interesting from the viewpoint of the number of possible singly-excited $2p^{-1}3d^1$ configurations as this number is quite small (Table~\ref{tab:objects}). 
	The nickel complex, possessing an almost filled $d$-shell, also features the small number of relevant electronic configurations similar to Ti, but its \ac{SOC} constant is larger by about a factor of three.
	The chromium complex has a more intricate electronic structure with the $d^3$ ground state, resulting in lots of excited states similar to the reference iron compound. 
	A standalone compound, in some sense, is \FeCO. 
	Due to the strong-field ligands, it has a singlet ground state, and the spin-state energetic pattern is substantially different from the other high-spin complexes.

\section{Computational details} \label{sec:comp_details}

	\begin{table}
		\caption{\label{tab:pulses} Summary of the pulse characteristics, see Eq.~\eqref{eq:pulse}.}
		\begin{ruledtabular}
			\begin{tabular}{lccc}
				
				Compound       & $\sigma$, fs & $\hbar\Omega$, eV & A, a.u. \\	
				\hline
				\TiO           & 0.2 & 470 & 1.5 \\
				\CrWat         & 0.2 & 588 & 2.5 \\
				\FeWat         & 0.2 & 716 & 6.0 \\
				\FeCO          & 0.2 & 728 & 6.0 \\
				\NiWat         & 0.2 & 875 & 9.0  
			\end{tabular}
		\end{ruledtabular}
	\end{table}

	All structures were optimized at the DFT level with the BLYP functional and cc-pVTZ basis set in \texttt{Gaussian} program package.~\cite{Frisch2009}
	A \ac{CASPT2}~\cite{Finley_CPL_1998,Forsberg_CPL_1997} geometry optimization was performed for some of the structures (see Table~\ref{tab:objects}) that led to bond shortening, slight changes in oscillator strengths and insignificant variations in dynamics. 
	Scalar relativistic effects were introduced via the Douglas-Kroll-Hess transformation \cite{Douglas_AP_1974} in conjunction with the all-electron ANO-RCC basis set~\cite{Roos_JPCA_2005} of VTZP quality. 
	The active space of 8 orbitals (three $2p$ and five $3d$) was found to give a good approximation~\cite{Bokarev_WCMS_2020} and is used for all species except for the \FeCO.
	Full \ac{CI} has been done for the $3d$ subspace (RAS2), while for the $2p$ subspace (RAS1), only one hole has been allowed. 
	For \FeCO, the $3d\sigma$ ($a_1^{\prime}$), four $3d$ ($e^\prime$ and $e^{\prime\prime}$), and $3d\sigma^\ast$ ($a_1^{\prime\ast}$) orbitals were added to the RAS2 as well as four $\pi^{\ast}$ orbitals to the RAS3 with only one electron allowed, resulting in 13 orbitals in the active space.
	The \ac{XAS} was calculated at the \ac{RASSCF} level of theory. 
	Note that all the experimental spectra were shifted to be aligned with calculations for the computational consistency as opposed to the conventional way of doing vice versa.

	The particular construction of basis functions and respective matrices in Eq.~\eqref{eq:hamiltonian} has been done as follows.
	First, molecular orbitals were optimized in a state-averaged \ac{RASSCF} procedure,~\cite{Malmqvist_JPC_1990} where averaging over all possible electron configurations has been performed.
	These orbitals were kept frozen during the propagation.
	In order to include dynamic correlation, the \ac{CASPT2} and \ac{RASPT2} \cite{Finley_CPL_1998,Forsberg_CPL_1997} methods has been used in specific cases.
	The \ac{SOC} coupling matrix $\mathbf{V}_{\rm SOC}$ is computed by means of the state interaction approach,~\cite{Malmqvist_IJQC_1986, Malmqvist_CPL_1989} implementing the atomic mean field integrals \cite{Schimmelpfennig1996,Marian_RCC_2001} method.
	It has proven itself to be a versatile tool for computing the L$_{2,3}$-edge absorption spectra calculations of transition metal complexes.~\cite{Josefsson_JPCL_2012,Wernet_JPCL_2012,Bokarev_PRL_2013, Bokarev_WCMS_2020} 
	The respective calculations have been done with \texttt{OpenMOLCAS} program package.~\cite{FernandezGalvan_JCTC_2019} 
	
	The width of the light pulse $\sigma$ was set to  0.2 fs in the time domain, see Eq.~\eqref{eq:pulse}, for all simulations.
	The carrier frequency $\Omega$ was chosen to correspond to the center between the L$_3$ and L$_2$ bands, see Table ~\ref{tab:pulses}. 
	The amplitude $A$ of the pulse was adjusted to ensure approximately the same depletion of the ground state in all simulations.
	The initial population of near-degenerate ground states in the high-spin complexes corresponded to the temperature of $T=300$\,K.

\section{Results} \label{sec:results}
	
	\subsection{Influence of ligand strength} \label{subsec:ligands}
	
		\begin{figure}
			\includegraphics[width=1\linewidth]{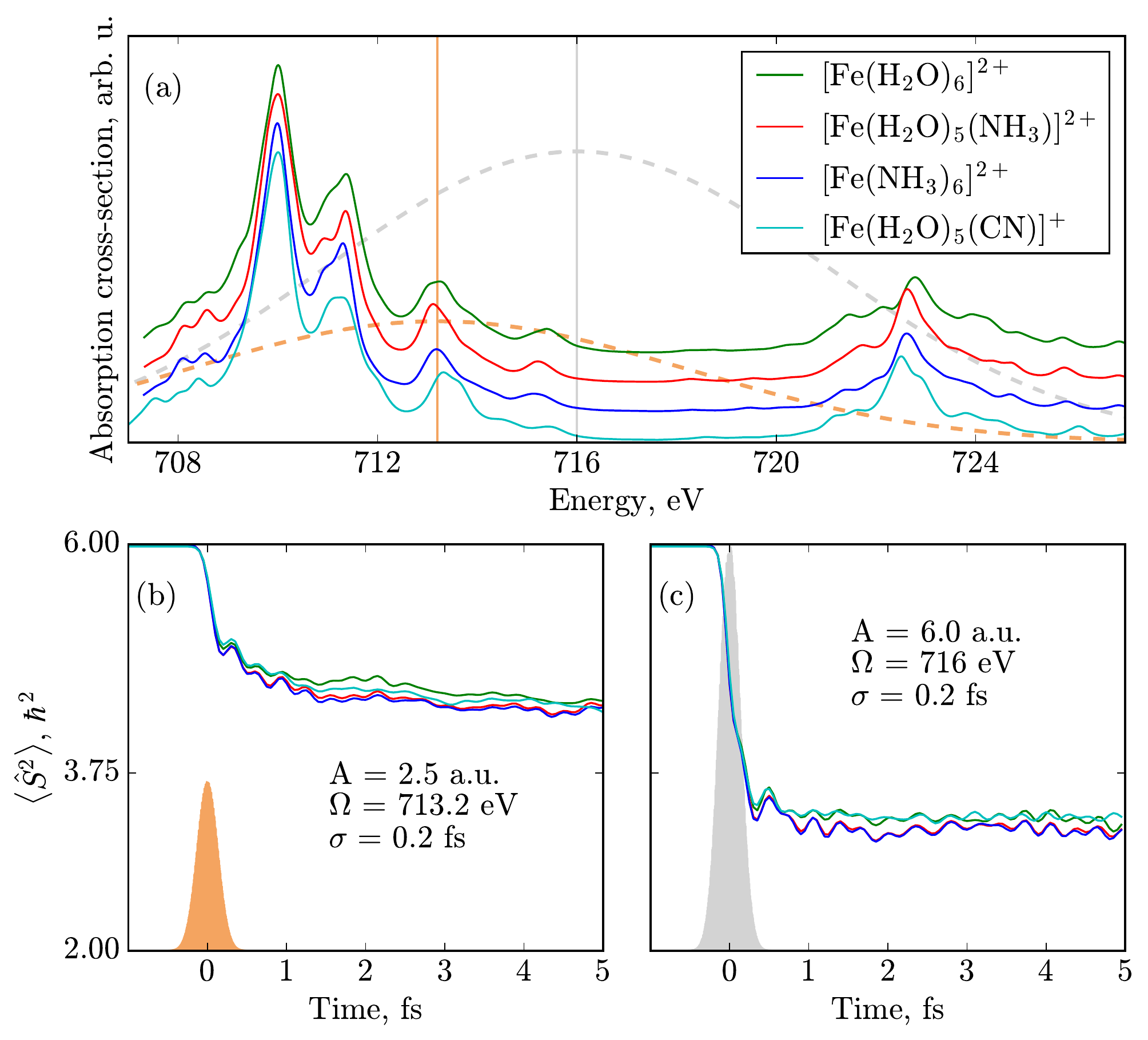}
			\caption{\label{fig:ligands} 
				(a) Calculated \ac{XAS} of the reference complex \FeWat and complexes from \textit{set 1}, see Table~\ref{tab:objects}. 
				(b) and (c) Time evolution of the $\langle \hat S^2 \rangle$ for these complexes for two different pulses; their characteristics are given in the respective panels, and filled curves depict the time envelopes. 
				The centers of the corresponding excitation bands in the frequency domain are also depicted in panel (a) with two vertical lines. 
				The amplitude has been selected to give a comparable depletion of the ground state.}
		\end{figure}

		The influence of the ligand surrounding on the dynamics has been studied on the example of Fe$^{2+}$ complexes with H$_2$O, NH$_3$ and CN$^{-}$ ligands.
		Despite different positions in the spectrochemical series (especially that of H$_2$O and CN$^{-}$), all these complexes have a quintet ground state, see Table~\ref{tab:objects}.
		\ac{XAS} for all members of the set represent dipole-allowed transitions from the $2p_{3/2}$ and $2p_{1/2}$ orbitals to the non-bonding $3d (t_{2g})$ and anti-bonding $3d\sigma^\ast (e_g)$ levels.
		Although one sees clear differences in the nature and energy of individual transitions between complexes, e.g., in the extent of spin-mixing, these differences are washed out upon lifetime broadening.
		Therefore, the \ac{XAS} spectra of different species are fairly similar, showing only minor differences in the L$_3$/L$_2$ energy splitting as well as in the structure of the L$_3$ edge, see Fig.~\ref{fig:ligands}(a).

		The spin-flip dynamics occurring in these complexes upon light excitation is illustrated in Fig.~\ref{fig:ligands}(b) and (c),
		where the time-dependent values of \Ssq\ are presented. 
		One can see that upon excitation, the expectation value \Ssq\ first quickly drops from the value of $6\hbar^2$, corresponding to the quintet state manifold, and then slower evolves after the pulse is over, exhibiting some oscillations.
		However, the final \Ssq\ does not reach the triplet value of $2\hbar^2$, evidencing a notable contribution from quintet states in the superposition.
			
		Concerning the influence of ligands, the same statement as for \ac{XAS} can also be made for the  dynamics. 
		It can be seen from a similar form of the respective $\langle \hat S^2 \rangle$ curves as a function of time in Fig.~\ref{fig:ligands}(b) and (c) for two different pulses.
		These pulses have different carrier frequencies and amplitudes and thus involve different groups of states in the dynamics.
		For instance, the gray pulse (panel (c)), centered between L$_3$ and L$_2$, overlaps with the latter edge in energy, whereas the orange pulse (panel (b)) barely touches it.
		This fact explains the larger yield of triplet states in the case of the gray pulse.
		Comparing the curves for different ligands, one can conclude that, at least for short pulses (broad in energy), the smearing of the fine details of the electronic structure occurs, leveling the differences due to ligands.
		Summarizing, the relatively small changes in the chemical environment, which do not lead to the change in the spin of the ground state or qualitative differences in the order of electronic states, cannot be used to tweak the character of the spin-flip dynamics.
		
		To consider a qualitatively different case, we now address the results for \FeCO.
		\begin{figure}
			\includegraphics[width=1\linewidth]{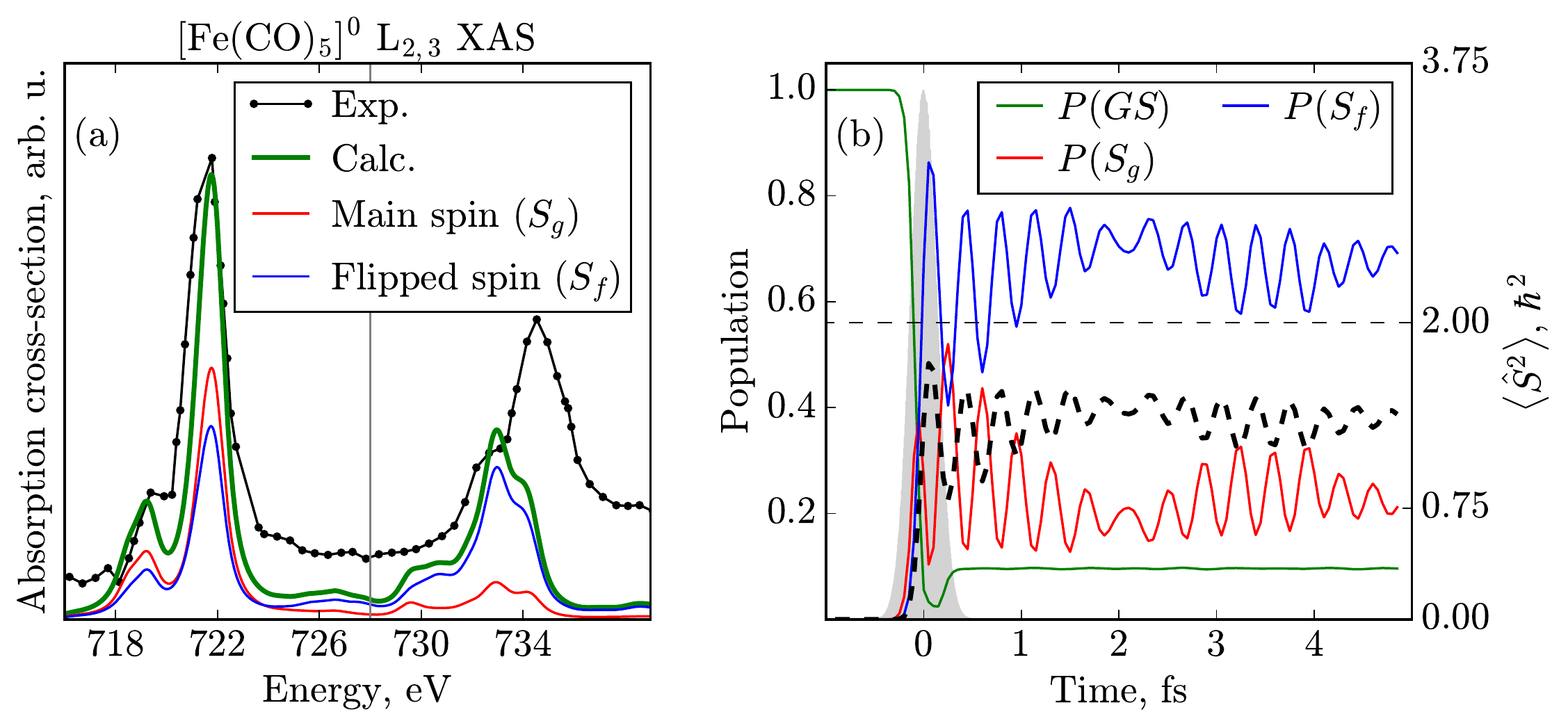}
			\caption{\label{fig:feco5} Results of modeling for the \FeCO\ complex: (a) Experimental (black dotted line) and calculated (green line) \ac{XAS}.
			The total intensity is decomposed according to the fraction of singlet (red curve) and triplet (blue curve) character of the respective states, see text. 
			(b) Evolution of the population of singlet $S_g$ = 0 (red) and triplet $S_f$ = 1 (blue) \ac{SF} states initiated by the pulse with characteristics given in Table~\ref{tab:pulses}. 
			The dashed line shows the expectation value of the $\hat S^2$ operator. 
			The value of $2\hbar^2$, marked with a horizontal line, corresponds to the pure triplet $S_f = 1$.}
		\end{figure}
		For this complex, all ligands have a strong field, and, in contrast to \FeCN, this results in the low-spin singlet ground state.
		\FeCO\ spectrum is less consistent with the experiment~\cite{Suljoti_ACIE_2013} than that of the \FeWat\ species (see Fig.~\ref{fig:feco5}(a)) as the \ac{SOC} splitting is underestimated by 2-3\,eV, but main spectral features can be clearly recognized.
		The lower intensity pre-peak at about 720\,eV is due to transitions to the $3d\sigma^\ast$ ($a_1^{\prime\ast}$) orbital and is thus somewhat similar to the transitions discussed before for the \textit{set 1}. 
		In contrast, the pronounced second peak of the L$_{3}$ edge at 722\,eV is a fingerprint of a strong $\pi$-backdonation~\cite{Suljoti_ACIE_2013} as it mainly corresponds to transitions from the $2p_{3/2}$ to the ligand $\pi^\ast$ orbitals, which are notably mixed with the iron $3d$ orbitals.
		Thus, excitation with the light pulse occurs from the ground singlet state to predominantly charge-transfer ones because of the larger transition strengths of the latter.
		The partial contributions of different spin states to the \ac{SOC}-coupled ones are illustrated in Fig.~\ref{fig:feco5}(a).
		The total intensity is partitioned according to the fraction of the singlet (red curve) and triplet (blue curve) \ac{SF} states contributions to the respective \ac{SOC}-state. 
		Essential for the current discussion is that the L$_3$ states have approximately equal contributions from singlet and triplets \ac{SF} states, whereas, for the L$_2$, triplets distinctly dominate.
			
		Dynamics in \FeCO\ is shown in Fig.~\ref{fig:feco5}(b) for the pulse centered between the L$_3$ and L$_2$ edges and overlapping with all dipole-allowed transitions.
		One sees the spin-flip from singlet to triplet happening much faster than the pulse duration, i.e., shortly after the initial singlet-singlet excitation (red curve).
		Remarkably, in contrast to other iron complexes from \textit{set 1}, the efficient spin transition occurs independent of the pulse characteristics.
		In this case, the final \Ssq\ is closer to the target triplet value of 2$\hbar^2$.
		Noteworthy, in the case of \FeCO\, pronounced oscillations in the state populations and the $\langle \hat S^2 \rangle$ are observed.	
		The time period of these oscillations (0.35\,fs) corresponds to the \ac{SOC}-splitting of 11.0\,eV.
		Naturally, when the symmetry is changed and the electronic structure is altered by the strong-field ligands and the dominant contributions from the charge-transfer states, the time-evolution changes qualitatively.
		The reasons for this fact will be further analyzed in Section~\ref{sec:discussion}.

	\subsection{Transition metal series} \label{subsec:transition_metals}

			\begin{figure*}
				\includegraphics[width=1\linewidth]{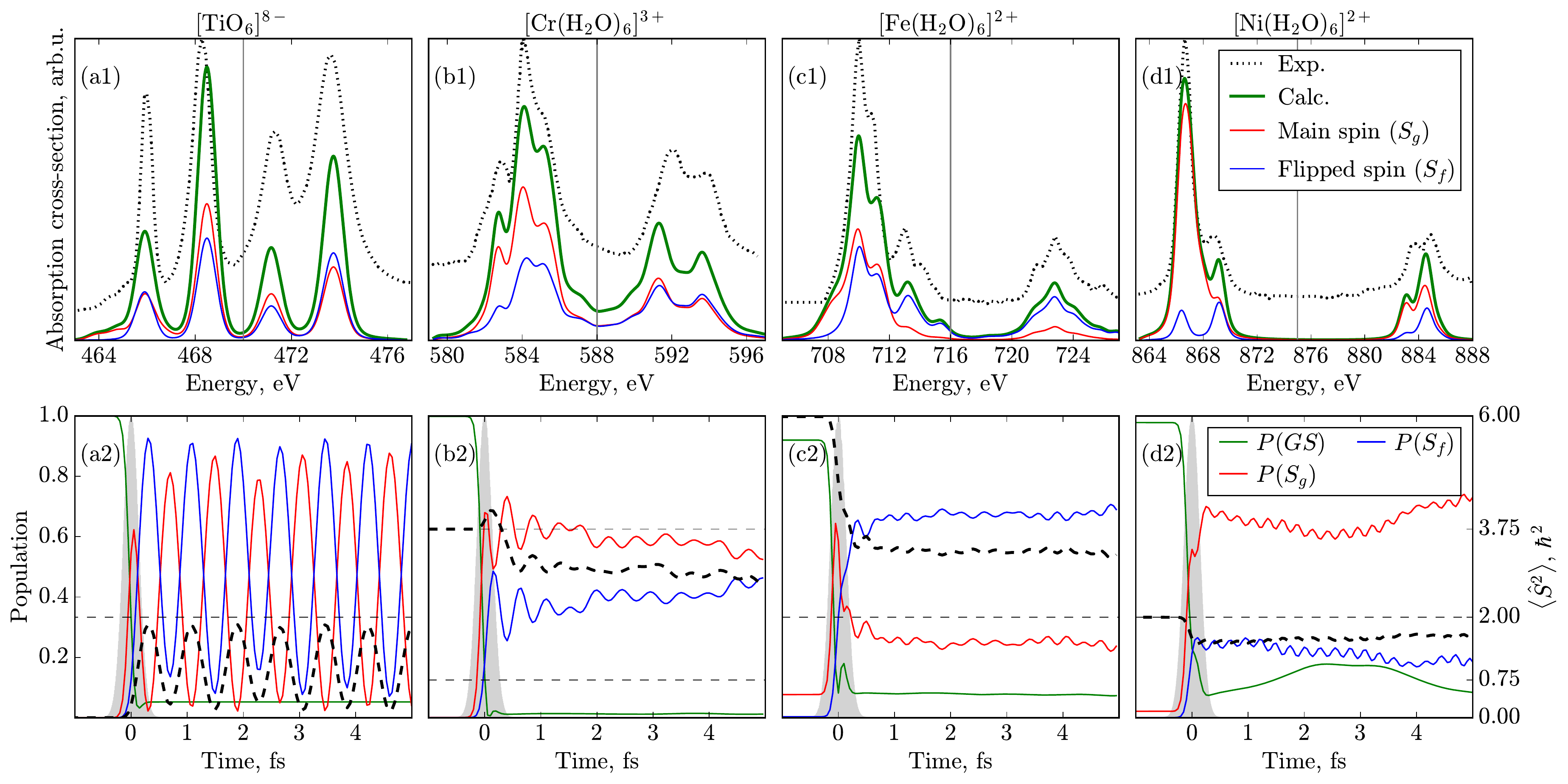}
				\caption{\label{fig:metals_spectra} 
				Upper row: the comparison of the experimental \ac{XAS} spectra~\cite{Woicik_PRB_2007,Wernet_JPCL_2012,Bokarev_PRL_2013,Josefsson_JPCL_2012}  with the calculated ones for species from \textit{set 2} (Table~\ref{tab:objects}). 
				Lower row: \ac{SF}-state population dynamics in the corresponding complexes initiated by pulses with characteristics given in Table~\ref{tab:pulses}. 
				All pulse amplitudes (light-gray) are normalized to the same height for the sake of clarity. 
				Partial populations (Eq.~\eqref{eq:populations}) of the ground states $GS$, excited states with the same spin $S_g$, and spin distinct by $\pm 1$ (flipped spin) $S_f$ are depicted in green, red and blue, respectively.
				Thick dashed lines give the actual $\langle \hat S^2 \rangle (t)$ for a particular complex. 
				Horizontal dashed lines indicate the expectation values $S(S+1)\hbar^2$ of the $\hat S^2$ operator for half-integer and integer spins $S$ relevant for each complex. 						
			}
		\end{figure*}
	
		The influence of the central atom is studied on the example of complexes from \textit{set 2}, see Table~\ref{tab:objects}.
		In the upper row of Fig.~\ref{fig:metals_spectra} the calculated L$_{2,3}$ absorption spectra are presented. 
		Overall, a reasonably good agreement with the experiment is reached already at the \ac{RASSCF} level of theory.
		For \CrWat\ spectra, the \ac{CASPT2} correction has been calculated, but it did not give any significant improvement of the agreement. 
		The \ac{SOC} splitting between the L$_3$ and L$_2$ bands, as well as the ligand-field splitting within L$_3$ band, is well reproduced for all systems under study. 
		The \TiO\ spectrum is compared to the Ti L$_{2,3}$-edge \ac{XAS} in SrTiO$_3$ \cite{Woicik_PRB_2007} as a reference for the $d^{0}$ system in the octahedral field of oxygen atoms. 
		It shows four clear peaks originating from the $2p^{-1}_{3/2}t_{2g}$, $2p^{-1}_{3/2}e_{g}$, $2p^{-1}_{1/2}t_{2g}$, $2p^{-1}_{1/2}e_{g}$ states. 
		A somewhat similar multiplet configuration can be roughly recognized for other species of the row, but one sees an intensity redistribution and appearance of additional peaks due to the stronger multiconfigurational character.
		
		The first row in Fig.~\ref{fig:metals_spectra} depicts the decomposition of spectra in the spin multiplicity of final states (red and blue curves). 
		Note that almost everywhere, the contributions from the states with ``main'' or ``ground'' spin $S_g$ (red) are higher than from the ones with the ``flipped'' spin $S_f$ (blue). 
		The only exceptions are the L$_2$ band of \TiO and the 712-727\,eV region of \FeWat\ shown in Fig.~\ref{fig:metals_spectra} (a1) and (c1), respectively. 
		
	 	The spin dynamics of the \textit{set 2} are shown in the lower row of Fig.~\ref{fig:metals_spectra}. 
	 	The \TiO\ compound features strong oscillations between the ``main spin'' and ``flipped spin'' states with a period of about 0.6\,fs, which corresponds to 6.9\,eV energy (Fig.~\ref{fig:metals_spectra}(a2)), correlating with the L$_3$/L$_2$ \ac{SOC} splitting.
	 	It is caused by the relatively small number of states involved in the dynamics giving rise to the Rabi-like oscillations.
		Other complexes demonstrate the same trend, i.e. oscillations have the characteristic period inversely proportional to the value of \ac{SO} multiplet energy separation $\Delta E [eV] = 4.14 / t [fs]$. 
		Along with the shortening of the oscillation period, their amplitude decreases in the row due to the increase of the \ac{SOC} energy separation. 
		In cases of chromium (Fig.~\ref{fig:metals_spectra}(b2)) and iron (Fig.~\ref{fig:metals_spectra}(c2)), the oscillations are dumped slowly which is related to a huge number of involved states. 
		Similar behavior has also been observed for the \FeCO\ (Fig.~\ref{fig:feco5}).
		
		The prevailing population of states with the ``flipped spin'' (blue line in the lower row of panels in Fig.~\ref{fig:metals_spectra}) is observed only in \TiO\ and \FeWat\ complexes, whereas for \CrWat\ and \NiWat\ the spin of the ground state (red line) stays dominant. 
		This behavior is also observed for the $\langle \hat S^2 \rangle$ curves (dashed line).
		For the former two cases, the \Ssq\ significantly deviates from the initial value and tends to the flipped value, namely singlet ($0\hbar^2$) to triplet ($2\hbar^2$) transition for \TiO\ and quintet ($6\hbar^2$) to triplet ($2\hbar^2$) for \FeWat.
		In turn, only a moderate spin transition can be seen for \CrWat\ (b2) and \NiWat\ (d2). 
		To summarize, there is no correlation between the efficiency of spin transition and the value of \ac{SOC}, which is opposite to what can be anticipated from general considerations.

\begin{figure}
			\includegraphics[width=1\linewidth]{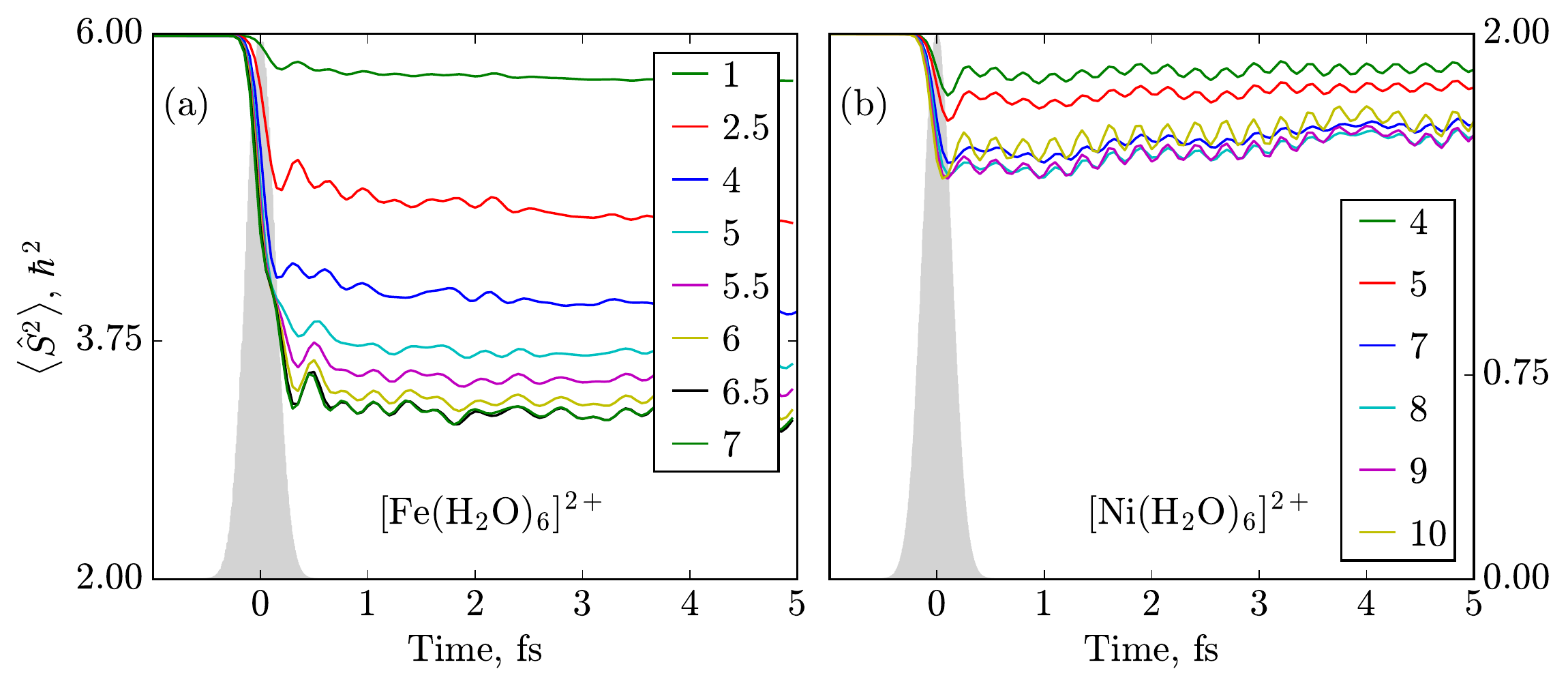}
			\caption{\label{fig:amplitudes} The dependence of $\langle \hat{S}^2 \rangle (t)$ on the amplitude $A$ of the incoming pulse, which is given in atomic units in the legend,  exemplified for (a) $\hbar\Omega=716$\,eV and $\sigma = 0.2$\,fs for \FeWat, and (b) $\hbar\Omega=875$\,eV and $\sigma = 0.2$\,fs for \NiWat.} 
		\end{figure}

		To strengthen this conclusion, one has to exclude the influence of the light pulse because the	amplitude has been selected differently for different complexes.
		It seems natural that the pulse strength notably influences the $\langle \hat{S}^2 \rangle$ for the same system, and one might argue that substantially increasing the amplitude one could achieve a more efficient spin conversion.
		This fact is illustrated in Fig.~\ref{fig:amplitudes}, where the dependence of $\langle \hat{S}^2 \rangle (t)$ on the amplitude of the pulse ranging from 1 to 7 a.u for iron (a) and from 4 to 10 a.u. for nickel (b) is shown.
		One can see that already at 6 a.u. for iron and 7 a.u. for nickel, there is no further increase in the yield of spin transition.
		Observed saturation takes place when the ground state population is almost completely depleted.
		These values of amplitude, at which the total population of the ground states $GS$ drops below 0.1, were chosen for all the complexes to exclude the influence of the pulse strength possibly.
		In transition metal row, the characteristic $A$ value increases as we excite in the center of L$_{2,3}$-edge, meaning that stronger pulse is needed to overlap with more energetically distant transitions separated by \ac{SOC} efficiently.
		As an example, note in Fig.\ref{fig:amplitudes}(a) the growing prominence of oscillations with the increasing amplitude, witnessing an involvement of more distant L$_3$/L$_2$ groups of states.
		
		Finally, the $3p \rightarrow 3d$ excitations have been considered.
		The respective \ac{XAS} spectra are given in \Supp.
		Although \ac{SOC} is also notable in this case (\ac{SOC} splittings are up to 6\,eV in \NiWat), the $S_g$ states are by far prevailing among the bright states and no spin-flip dynamics is observed.
		That is why this case will be not further discussed here.
		
\section{Discussion} \label{sec:discussion}
			 	
	 \begin{figure}
	 	\includegraphics[width=\linewidth]{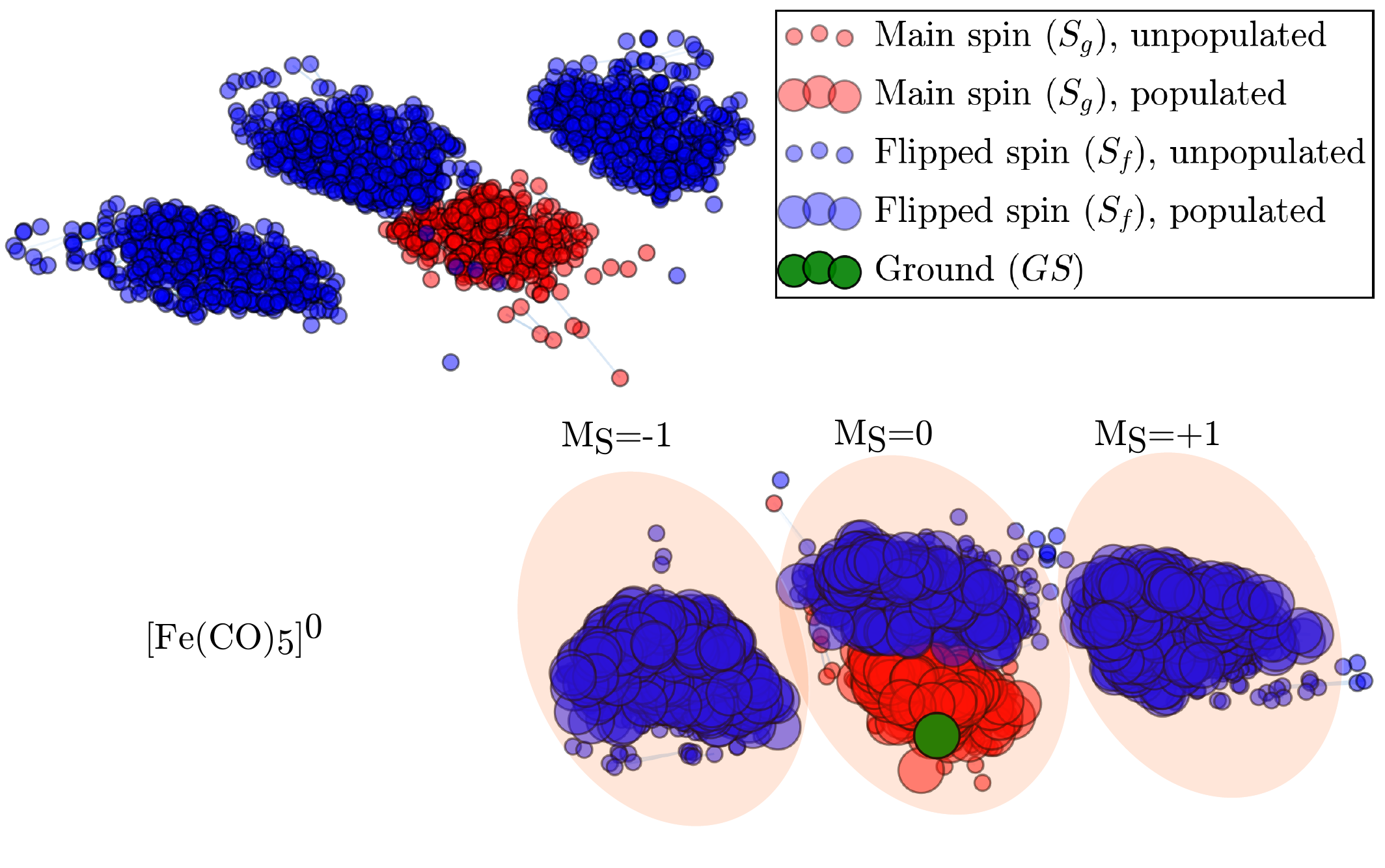}
	 	\caption{\label{fig:states_feco5} 
	 		Force-directed graph, showing the clustering of the states of \FeCO\ according to the transition-dipole and \ac{SOC} coupling, see Eq.~\ref{eq:spring_force}.   
	 		Each node corresponds to one of the 3766 \ac{SF} basis states $\Phi_{i}^{(S,M_S)}$:
	 		green -- $GS$ singlet states, 
	 		red -- excited singlet states, 
	 		blue -- excited triplet states.  
	 		Large circles indicate states participating in the dynamics, i.e., having a notable maximal population; small circles correspond to the ``spectator'' states acquiring no population.}
	 \end{figure}
 	
 	\begin{figure*}
 		\includegraphics[width=1\linewidth]{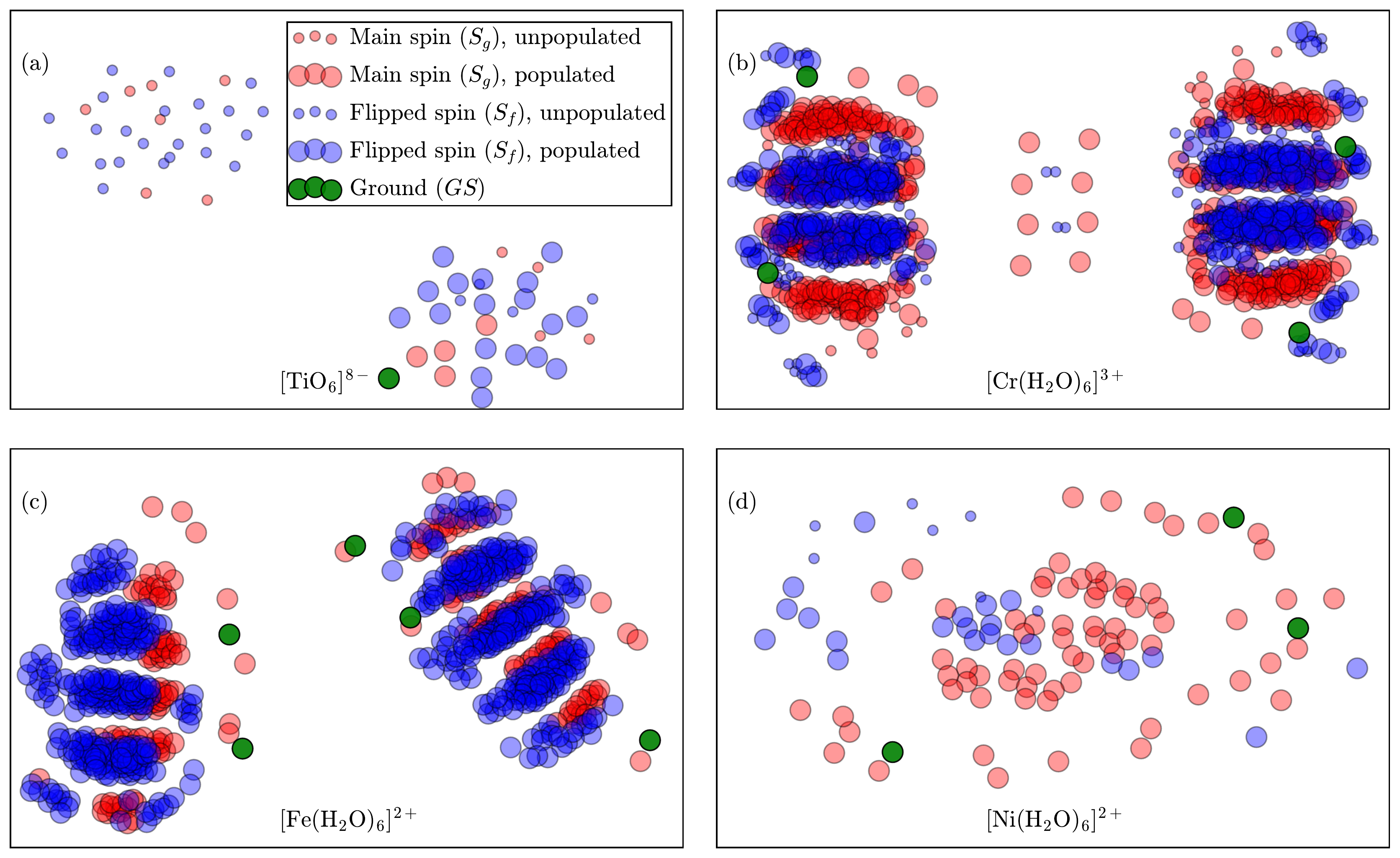} 		
 		\caption{\label{fig:states_tm_row} Clustering of states for the complexes of \textit{set 2}, see caption of Fig.~\ref{fig:states_feco5}.}
 	\end{figure*}
 	%\rt{INFLUENCE OF TEMPERATURE ON DYNAMICS!!!}
 	
 	From the discussion in Sections~\ref{subsec:ligands} and \ref{subsec:transition_metals} one can make three general observations.
 	First, the qualitative character of the dynamics is only barely dependent on the chemical nature of ligands unless the electronic structure is altered completely.
 	The examples are high-spin complexes of the \textit{set 1}, demonstrating a very similar behavior, and \FeCO, possessing the low-spin ground state and exhibiting a completely different energetic pattern of spin-states.
 	Second, in contrast to expectations, the value of \ac{SOC} splitting does not play a decisive role in the character of dynamics, as seen from the comparison of different metals.
 	For instance, \NiWat\ has the largest \ac{SOC} constant in the considered series but does not show prominent spin-flip dynamics.
 	On the other hand, \TiO\, with its \ac{SOC} constant being by a factor of three smaller, demonstrates intricate dynamics.
 	Third, the critical point is the ratio between the \ac{SF} states with different spins constituting the \ac{SOC}-eigenstates. 
 	It can be seen from the \ac{XAS} decomposition into $S_g$ and $S_f$ contributions in Figs.~\ref{fig:feco5} and \ref{fig:metals_spectra}. 
 	Indeed, a significant spin-flip was observed for titanium and iron compounds, where bright states with the prevailing amount of the $S_f$ spin contributions are dominating in \ac{XAS} for the high-energy flank of the L$_3$ and the whole L$_2$ edges. 
 	It means that at specific energy ranges more \ac{SF} ``flipped'' states can be accessed by the excitation.
  	For the $3p$ excitation, there are no such ranges, and spin-flip is not observed.
 	However, a profound analysis going beyond these simple observations is complicated due to the vast amount of the electronic states which are coupled in an entangled way.
 	
 	To shed light on the reasons for such behavior and attain a more mechanistic understanding, let us consider a somewhat simplified model, see Fig.~\ref{fig:scheme}(b).
 	Assuming that we are working in the saturated regime (Fig.~\ref{fig:amplitudes}) and thus can  neglect the details of the incoming light pulse for simplicity, the dynamics are governed by two factors -- strengths of the dipole transition and that of the \ac{SOC}.
 	Let us follow the density matrix evolution and namely, its diagonal elements in the basis of \ac{SF} states.
 	Initially, an entire population resides in the ground state and the lowest excited states, which are populated according to the respective Boltzmann factors, see Section~\ref{sec:comp_details}.
 	The light pulse couples these initial states with the core ones through the respective transition dipole ${\mathbf{d}}$ matrix elements.
 	Note that the spin quantum number is conserved $GS \rightarrow S_g$ due to the spin selection rules. 
 	Reflecting this fact, in Fig.~\ref{fig:metals_spectra} the red line $S_g$ rises simultaneously with the arriving pulse.
 	In the \ac{SOC} picture, states with strictly defined spin do not exist, as the spin quantum number is not conserved; thus, the predominant population of the $S_g$ \ac{SF} states corresponds to a non-stationary superposition of \ac{SOC} eigenstates. 
 	After the initial population of the core-excited $S_g$ states, all \ac{SF} states get mixed regardless of their spin through $\mathbf{V}_{\rm SOC}$. 
 	That is why the blue $S_f$ line goes up parallel to the red one but after a short delay (see Fig.~\ref{fig:metals_spectra}).
 	Once the pulse is switched off, the dynamics are governed solely by the elements of $\mathbf{V}_{\rm SOC}$.
 	This free dynamics is then determined by the populations accumulated during the pulse in the bright core-excited states with spin $S_g$ and their \ac{SO} coupling to other states with both $S_g$ and $S_f$.
 	%These states can be called \rt{``relevant''} states participating in ultrafast dynamics.
 	 
 	Provided a large number of coupled states, we employ a concept widely used in the big data analysis to illustrate some trends. 
 	Let us start the discussion from the example of \FeCO\ complex; see Fig.~\ref{fig:states_feco5}.
  	This figure is obtained with the NetworkX package \cite{Hagberg_Pot7PiSC_2008} implementing the force-directed graph drawing algorithm by Fruchterman and Reingold.~\cite{Fruchterman_SPE_1991} 
 	Here, each node corresponds to one of the 3766 \ac{SF} basis states $\Phi_{i}^{(S,M_S)}$ and the color encodes their nature, e.g., ground as well as excited states with spins $S_g$ and $S_f$.
 	The size of the nodes, in turn, denotes whether the state is involved in dynamics (we call it participating) or stays mainly unpopulated (spectator).
 	The distances between nodes are optimized to minimize spring-like forces between them.
 	If the pulse characteristics are left besides the discussion, the force ($F_{ij}=-k_{ij}\Delta x_{ij}$) between nodes $i$ and $j$ corresponds to the spring constant 
 	\begin{equation}\label{eq:spring_force}
 		k_{ij}=c|({V}_{\rm SOC})_{ij}| + |d_{ij}|\, ,
 	\end{equation} 
 	where $c$ is a factor governing the relative importance of the two couplings. 
 	It has been adjusted for visual clarity to illustrate the clustering of states.
 	These two quantities in the sum are correlated to the degree of spin conversion. 
 	The dipole matrix identifies states which can be directly populated by the light absorption from the initial state manifold (which is denoted as green nodes in the figure).
 	The subsequent dynamics is governed mainly by the strength of \ac{SOC}.
 	The combination of these two quantities allows considering the entangled effects together.
 	
 	Looking at Fig.~\ref{fig:states_feco5}, one can notice the following peculiarities.
	According to the above criteria, the states group in two main clusters which are separated from each other and thus are connected neither by transition dipole nor by \ac{SOC}.
	Inside both clusters, one can distinguish four subgroups.
	The red one corresponds to the singlet ($S_g$) excited states ($M_S=0$); the three blue subgroups are triplet ($S_f$) states grouped by their $M_S$ quantum number.
	Inside of the smaller subgroups, both $\mathbf{d}$ and $\mathbf{V}_{\rm SOC}$ contributions keep nodes together, whereas between them only the \ac{SO} interaction is non-zero.
	This is due to the spin selection rules ($\Delta S=0, \Delta M_S = 0$) for the dipole transitions, causing the blocked structure of $\mathbf{d}$.
	Remarkably, the participating states (big nodes) are found only in the cluster, where the single ground state (green node) is entering.
	Thus, the second big cluster is completely excluded from the dynamics.
	Even in the former cluster, a relatively small amount of states (about 200 out of 1800) are populated during dynamics.
	The last important notice is that the amount of the ``spin-flipped'' states in this participating cluster is larger than that of the spin-conserved ones.

	The graphs for the other compounds of \textit{set 2} and the reference \FeWat complex are presented in Fig.~\ref{fig:states_tm_row}.
	For most of the species (apart from \NiWat), the states also group in two major clusters.
	\TiO\ (panel (a)), however, does not show subdivision according to the $M_S$ quantum number for the triplet states.
	For this complex, the overall number of states is the lowest among all systems.
	The singlet ground state enters only one cluster similar to \FeCO.
	Analogously to the latter, the amount of triplet $S_f$ states is dominating over the $S_g$ singlet states.
	\CrWat and \FeWat systems shown in panels (b) and (c), respectively, demonstrate similar clustering.
	In these two cases, the $M_S$-components of the ground state are distributed between two major clusters, thus leading to the involvement of both groups of states into the dynamics.
	Therefore, almost all considered states are populated within the first femtosecond.
	This behavior should, however, depend on the temperature: for low temperatures, only one component of the ground state may be initially populated.
	Both clusters show a distinct splitting according to the $M_S$ quantum number.
	The difference between the two systems is the ratio between numbers of $S_f$ and $S_g$ states.
	The flipped states are prevailing in the case of \FeWat and represent a minority in the case of \CrWat.
	\NiWat is somewhat similar to \TiO\ since the total number of states is quite small.
	The three components of the ground state are also uniformly distributed throughout the cluster.
	The overall $M_S$ grouping is less pronounced but still can be seen in the central part of the panel (d).
	In contrast to other cases, the separation of states into two clusters is not present.
	One should also note the dominating number of the $S_g$ states.
	
	Although the graphs given in Figs.~\ref{fig:states_feco5} and \ref{fig:states_tm_row} provide a convenient visualization of the connections between different states, they do not allow to make an unambiguous conclusion about the decisive factors, which could be used for the \textit{a priori} assertion on the efficiency of spin flip for an arbitrary system.   
	The only factor which seems to favor the efficient transition is the dominating number of spin-flipped states over the states with the ground state spin.
	Such a situation is observed for \FeCO, \FeWat, and \TiO, being efficient systems, and is not observed for \CrWat and \NiWat, showing no prominent spin transition.
	As described for the case of \textit{set 2}, this domination can be present in some energy ranges and be absent for the other.
	This fact, explains the dependence of the efficiency on the particular pulse characteristics used for the excitation, see Refs.~\citenum{Wang_PRL_2017, Wang_PRA_2018}. 
	In this respect, the proper pulse leading to spin transition should necessarily overlap with the spectral regions, where $S_f$ states dominate.

\section{Conclusions} \label{sec:conclusions}
	
	This article represents an extension of the previous study of the ultrafast spin-flip dynamics in the core-excited states, which has been performed for a prototypical Fe$^{2+}$ complex.~\cite{Wang_PRL_2017, Wang_PRA_2018}
	There, the occurrence of the spin transition within the time window of hundreds of attoseconds has been observed, being also dependent on the characteristics of the exciting X-ray light pulse.
	Here, we address the main question, what is the crucial factor, influencing the spin dynamics in terms of the yield of the spin-flipped states? 
	For example, how central metal ion and surrounding ligands influence the extent of the transition.
	
	An intuitive answer can be suggested based on the two-level model, where the probability of the transition between states is proportional to the square of the coupling matrix element.~\cite{May2011}
	In particular, the efficiency of spin-flip should be proportional to the \ac{SOC} constant. 
	Therefore, one expects the population transfer from states with the spin of the ground state ${S_g}$ to ones with a different spin ${S_f}$ to increase from left to right in the periodic table.
	However, the situation appears to be more complicated. 
	Although the values of the \ac{SOC} matrix $\mathbf{V}_{\rm SOC}$ responsible for L$_{2,3}$-splitting are indeed important, the number of the relevant states plays a decisive role.
	For instance, the \ac{SOC} strength in \NiWat\ is three times larger than in \TiO, but the small number of the accessible spin-flipped ${S_f}$ states makes the whole process inefficient in the former case, while in the latter, the spin dynamics is much more prominent.
	
	Importantly, the exciting pulse should overlap with the spectral regions where $S_f$ states dominate. 
	Here, the decomposition of \ac{XAS} provides a hint about how many states of different multiplicities are presented and what is the chance to have enough relevant states in order to observe a target effect. 
	Relevant states are those that are coupled to the ground states by the dipole matrix elements either directly or indirectly through \ac{SOC}. 
	%\rt{Ligand-field splitting in this sense provides connection between states of the same group.}
	
	%\bt{Briefly how objects have been chosen.}

	The effect seems to be stable to moderate changes in the coordination sphere.
	For instance, the exchange of ligands situated close to each other in the spectrochemical series (e.g., H$_2$O and NH$_3$) does not lead to the qualitative changes in the rate and completeness of the spin dynamics. 
	However, ligands can substantially change the electronic structure of the outer valence shell, altering the relative energetic stability of the spin states.
	It is observed for \FeCO, where the ground state spin is changed to a singlet in contrast to \FeWat with its quintet ground state.

	In conclusion, the character and efficiency of the dynamics have to be analyzed on a case-to-case basis, as no general trends have been observed.
	We note that a crucial issue for the spin-flip yield is the rate of electronic and nuclear dephasing.
	On the example of \TiO\ and \FeCO\ systems, one observes the oscillations of \Ssq\ with time, and the final spin should strongly depend on when the oscillations will be effectively damped due to decoherence.
	However, this issue requires further investigation.
	
%	\bt{What can it be useful for? Firstly, it is interesting from the theoretical point of view as a process, which is complementary to the core-hole decay. Since different states have different decay channels multiconfigurational systems with very dense spectrum have to show very interesting interplay.  
%	Thus, future investigations will be able to shed the light on decay times and in conjunction with Auger rate calculations give quantitative estimations.
%	Another point is the possible applications in spintronics. One can speculate on either memory storage or some logical gates operating in the attosecond domain.}
	
\begin{acknowledgments}
	Financial support from the Deutsche Forschungsgemeinschaft Grant No. BO 4915/1-1 (V.K. and S.I.B.) and from the National Natural Science Foundation of China Grant No. 11904215 (H.W.) is gratefully acknowledged.
\end{acknowledgments}	

\bibliography{Spin_dynamics_ChemEnv_4}

\end{document}